\documentclass{article}
\usepackage{spconf,amsmath,graphicx}
\usepackage{multirow}
\usepackage[colorlinks=true,linkcolor=blue, citecolor=blue]{hyperref}
\usepackage{bbm, bm}
\usepackage{bbold}
\usepackage{tablefootnote}

\title{Asymmetric Clean Segments-Guided Self-Supervised Learning for Robust Speaker Verification}
\twoauthors
 {Chong-Xin Gan, Man-Wai Mak\thanks{This work was supported by the RGC of Hong Kong SAR, Grant No. PolyU 15210122, and the NSTC of Taiwan, Grant No. 112-2634-F-A49-006.} and Weiwei Lin }
	{Dept. of Electrical and Electronic Engineering \\
	The Hong Kong Polytechnic University \\
        Hong Kong SAR}
 {Jen-Tzung Chien}
	{Inst. of Electrical and Computer Engineering\\
	National Yang Ming Chiao Tung University \\
        Taiwan}
\begin{document}
%
\maketitle
\begin{abstract}
Contrastive self-supervised learning (CSL) for speaker verification (SV) has drawn increasing interest recently due to its ability to exploit unlabeled data. Performing data augmentation on raw waveforms, such as adding noise or reverberation, plays a pivotal role in achieving promising results in SV. Data augmentation, however, demands meticulous calibration to ensure intact speaker-specific information, which is difficult to achieve without speaker labels. To address this issue, we introduce a novel framework by incorporating clean and augmented segments into the contrastive training pipeline. The clean segments are repurposed to pair with noisy segments to form additional positive and negative pairs. Moreover, the contrastive loss is weighted to increase the difference between the clean and augmented embeddings of different speakers. Experimental results on Voxceleb1 suggest that the proposed framework can achieve a remarkable 19\% improvement over the conventional methods, and it surpasses many existing state-of-the-art techniques.

\end{abstract}
\section{Introduction}
\label{sec:intro}

Speaker verification (SV) aims to recognize if two utterances are spoken by the same person \cite{KINNUNEN201012, 9889705, mak2020machine}. This technique can be integrated into various applications, including financial security \cite{eskimez2018front} and forensic voice analysis \cite{4806209}. Earlier work on SV trained a Gaussian mixture model on a population of speakers and aligned acoustic frames against the Gaussians in the model to extract utterance-level features called the i-vector \cite{dehak2010front}. The i-vectors of a target speaker and an unknown speaker were fed into a backend classifier, such as probabilistic linear discriminant analysis (PLDA) \cite{prince2007probabilistic} for scoring. The classifiers effectively decoupled the speaker-specific information from potentially misleading channel factors. 

\begin{figure}[t]
    \centering
    \includegraphics[scale=0.31]{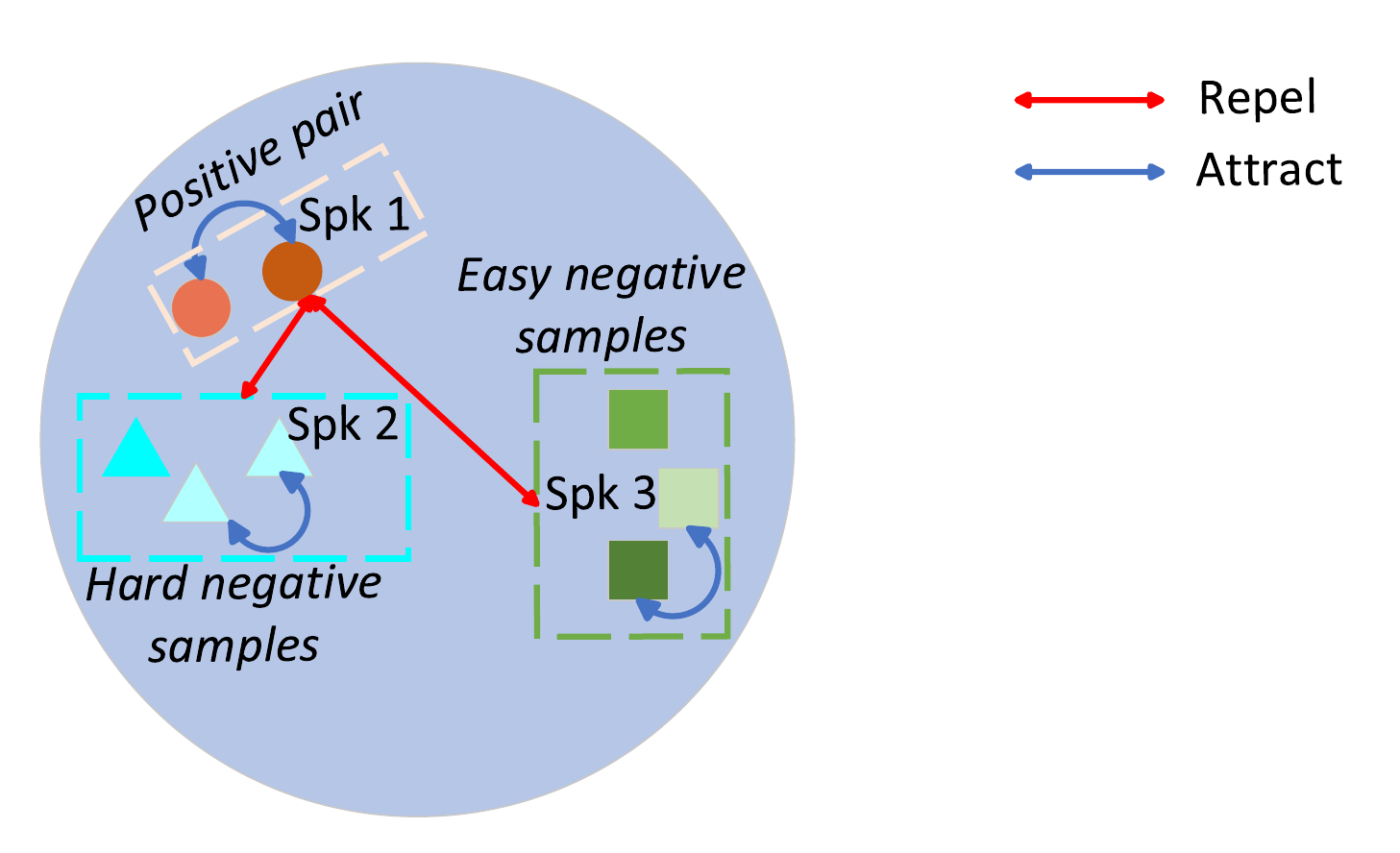}
    \caption{Distribution of speakers in the embedding space. Easy negative samples are easily differentiated from the target speaker (Spk 1), whereas hard negative samples are confusable with the target speaker. The hard negatives may mislead the network to converge towards a suboptimal solution.}
    \label{fig:enter-label}
\end{figure}

With the emerging trend of deep neural networks (DNNs), numerous studies \cite{snyder2018x, 7846260, desplanques2020ecapa, chung2019delving} demonstrated superior results by training networks with different deep architectures and loss functions \cite{wan2018generalized, liu19f_interspeech, huang2018angular, zhang2017end}. The significant accomplishments achieved so far heavily relied on a substantial amount of annotated data. Given the labeled data, a network can learn good speaker representations by minimizing the loss between the output probabilities and the ground truths. Labeled datasets, however, are difficult and expensive to collect. Thus, many researchers have shifted their focus to using self-supervised learning (SSL) for training speaker embedding networks. One category of SSL, which uses contrastive objectives, provides a practical solution. This category of approaches is inspired by some cutting-edge frameworks such as SimCLR \cite{chen2020simple} and MoCo \cite{he2020momentum}. These contrastive-based algorithms considered two non-overlapping segments from the same utterance as a positive pair and treated the segments from different utterances as negative pairs \cite{zhang2021contrastive, xia2021self, mun2020unsupervised, lepage2022label}. The network was optimized with the objective of pulling positive samples (embeddings from the same speaker) closer and negative samples (embeddings from different speakers) apart. 


\begin{figure*}[ht]
    \centering
    \includegraphics[scale=0.19]{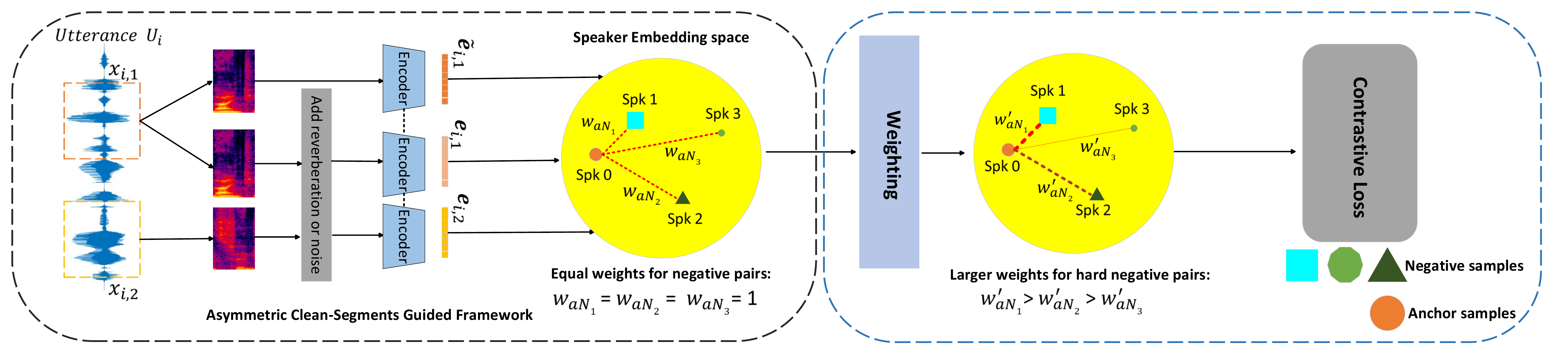}
    \caption{Overview of asymmetric clean segments-guided self-supervised learning framework. One additional clean segment is fed into the speaker encoder to form extra positive and negative pairs. The weighting mechanism guides the network to pay attention to hard negative pairs. The proposed contrastive loss is used to optimize the speaker encoder.}
    \label{fig:fig2}
\end{figure*}

Researchers commonly employ data augmentation techniques to reduce non-speaker information in the positive pairs, e.g., adding different types of noise to two segments of the same utterance. Adding noise and reverberation has become a standard practice to enhance the robustness of a speaker encoder. However, researchers also argued \cite{tian2020makes, xiao2021what} that not all types of augmentation are good for learning representations. In particular, excessive noise and reverberation may lead to the loss of speaker information. On the other hand, if the augmentation types are too alike, the model may be biased towards a few acoustic conditions, causing poor performance in unseen conditions. Therefore, preventing an embedding network from encoding undesirable factors is crucial. To this end, the authors of \cite{huh2020augmentation, tao2022self, li2020contrastive} used adversarial training to suppress the channel effects in the embeddings. However, training the channel discriminator and the speaker classifier in an adversarial manner is challenging. In this study, we propose a simple solution from a different perspective. Specifically, we introduce clean segments without augmentation into the training process to form extra positive and negative pairs to mitigate speaker information loss. 

Several studies \cite{robinson2021contrastive, NEURIPS2020_f7cade80} have proposed strategies to identify and utilize hard negative pairs to strengthen the robustness of encoders. Fig. \ref{fig:enter-label} shows some examples of hard negatives that are confusable with the target speaker. Paying more attention to these challenging negative pairs (i.e. the pairs formed by the target and the hard negatives) may help a speaker encoder learn robust representations with a large inter-class margin. However, this aspect has not been widely explored in the speech community. Recently, \cite{li2023discriminative} proposed a class-aware attention mechanism that weights the positive and negative pairs according to the similarity of the embeddings with their class representatives. This approach effectively handled hard negatives but relies on speaker labels, which may not be available. Motivated by the same goal, we compute the weights that are dependent on the similarity scores of negative pairs. These weights can be dynamically adjusted according to the similarity between the target and negative samples. The method achieves impressive performance on the task of using Voxceleb dataset.

In summary, we proposed integrating clean segments into the training pipeline to alleviate the loss of speaker-dependent information caused by data augmentation. Meanwhile, we introduced a weighting mechanism to enlarge the inter-class margin by actively paying attention to hard negative pairs. Experimental results showed that the proposed method can achieve superior results, outperforming many existing approaches.
\section{Methodology}
\label{sec:format}

To address the aforementioned challenges, we developed an \underline{\textbf{A}}symmetric \underline{\textbf{C}}lean \underline{\textbf{S}}egments-\underline{\textbf{G}}uided (ACSG) self-supervised contrastive learning framework for speaker verification. On top of the ACSG loss, we introduce the similarity-dependent weighting factors to weight across different negative pairs, which we refer to as W-ACSG. The weighting mechanism enhances the contribution of hard negative pairs in contrastive learning, making the speaker embedding network more robust. The architecture of the proposed method is illustrated in Fig.~\ref{fig:fig2}. To counteract any potential loss of speaker information caused by random augmentation, we feed additional clean segments alongside the two noisy segments into the speaker encoder. 

%

\subsection{Asymmetric Clean Segments-Guided Framework}

Given a dataset, we apply data augmentation (adding noise and reverberation) to its utterances. We consider the original speech segments from these utterances as clean and treat those corrupted by data augmentation as noisy. A mini-batch comprising $N$ utterances \{$U_{1}, \ldots, U_{N}$\} are assumed to be spoken by $N$ different speakers. This assumption is based on the low probability of false negative pairs, due to the small batch size and the large number of speakers in the dataset \cite{zhang2021contrastive}. Two non-overlapping segments, denoted as $x_{i, 1}$ and $x_{i, 2}$, are randomly truncated from the utterance $U_{i}$. After performing data augmentation, two noisy segments are fed into the speaker encoder $f(.)$ to obtain the speaker embeddings $\bm{e}_{i, 1}$ and $\bm{e}_{i, 2}$: 

\begin{equation}
    \bm{e}_{i, j} = f(x_{i, j}), \quad j=1, 2.
    \label{eq1}
\end{equation}

Segments from the same utterance should capture the same speaker information; thereby, they form the positive pairs. On the other hand, the segments from different utterances form the negative pairs. To pull the embeddings of positive pairs together while pushing the embeddings of negative pairs apart, we use the contrastive objective \cite{chen2020simple}:

\begin{equation}
    {\cal L}_{i,j}^{\mbox{\tiny{CSL}}} = - \log{\frac{\exp[{\cos{(\bm{e}_{i, 1}, \bm{e}_{i, 2}) / \tau}}]}{\sum_{k=1}^N \sum_{l=1}^2 \mathbb{1}_{k \neq i \atop
    j \neq l} \exp{[\cos{(\bm{e}_{i, j}, \bm{e}_{k, l})} / \tau]}}},
    \label{eq.2}
\end{equation}
where $i$ indexes the samples in a mini-batch, $N$ is the number of samples in the batch, and $\cos(\bm{a},\bm{b})=\frac{\bm{a}\cdot\bm{b}}{\|\bm{a}\|\|\bm{b}\|}$ is the cosine similarity. In Eq.~\ref{eq.2}, $\tau$ represents the contrastive temperature, which will be omitted in the following equations for simplicity, and the indicator function $\mathbb{1}$ outputs a $1$ when $k \neq i$ and $l \neq j$; otherwise it outputs a $0$.

Excessive augmentation could result in imperfect speaker representation because of the potential loss of speaker information in the augmentation process. We adopt a novel strategy to avoid such a situation. Specifically, we include the clean segments for training. To manage computational complexity, for each utterance, we only include the first clean segment in the training process. As a result, each utterance has three segments for contrastive learning: two noisy segments (augmented) and one clean (original) segment. For each mini-batch, $3 N$ segments are fed into the speaker encoder. For each utterance, we form an additional positive pair by considering the first clean segment and the second noisy segment to prevent the speaker encoder from learning the contextual information. The negative pairs are formed by clean and noisy segments sampled from different utterances. We reformulated the contrastive loss by a new Asymmetric Clean Segments-Guided (ACSG) contrastive loss: 

\begin{footnotesize}
\begin{equation}
    {\cal L}_{i,j}^{\mbox{\tiny{ACSG}}} = - \log{\frac{\exp{[\cos{(\bm{e}_{i, 1}, \bm{e}_{i,2})} + \cos{(\tilde{\bm{e}}_{i, 1}, \bm{e}_{i,2})}]}}{\sum_{k=1}^N \sum_{l=1}^{2} \mathbb{1}_{k \neq i \atop
    j \neq l} \exp{[\cos{(\bm{e}_{i, j}, \bm{e}_{k, l}) + \cos{(\tilde{\bm{e}}_{i, j}, \bm{e}_{k, l})}}]}}},
\end{equation}
\end{footnotesize}

\noindent where $\tilde{\bm{e}}_{i}$ denotes the embedding vector from the clean segment of $U_{i}$. The ACSG loss enhances the diversity of training samples by incorporating clean segments for network training. This approach efficiently utilizes the available samples.


\subsection{Similarity-Dependent Weights for Hard Negatives}

Negative pairs in contrastive learning generally can be divided into hard and easy negative pairs. Hard negative pairs refer to segments spoken by different speakers but still have high similarity scores. On the other hand, easy negative pairs exhibit significant dissimilarities between segments, enabling their embeddings to be easily distinguished. To tackle the challenge posed by hard negative pairs, we implement a weighting mechanism for the proposed ACSG contrastive loss.  

Contrastive learning is enhanced by emphasizing those confusing pairs with hard negative samples. Treating contrastive learning as single-sample discrimination, the weighting scheme is inspired by employing a large-margin classifier for a contrastive learning problem. As indicated in Eq.~\ref{eq.4} and Eq.~\ref{eq.5}, we compute two similarity-dependent weights, $\beta_{n}$ and $\beta_{c}$, to amplify the contribution of negative pairs when their similarity scores are high and reduce their importance when their similarity is low. Here, we describe the method for a single clean segment. However, our approach can be extended to multiple clean segments.

\begin{footnotesize}
\begin{equation}
\begin{split}
    &{\cal L}_{i,j}^{\mbox{\tiny{W-ACSG}}} = \\
    & - \log{\frac{\exp{[\cos{(\bm{e}_{i, 1}, \bm{e}_{i,2})} + \cos{(\tilde{\bm{e}}_{i, 1}, \bm{e}_{i,2})}]}}{\sum_{k=1}^N \sum_{l=1}^{2} \mathbb{1}_{k \neq i \atop
    j \neq l} \exp{[\beta_{n}\cos{(\bm{e}_{i, j}, \bm{e}_{k, l}) + \beta_{c}\cos{(\tilde{\bm{e}}_{i, j}, \bm{e}_{k, l})}}]}}},
\label{eq.4}
\end{split}
\end{equation}
\label{eq}
\end{footnotesize}
\noindent where
\begin{equation}
    \beta_{n} = \exp{(\cos{(\bm{e}_{i, j}, \bm{e}_{k, l})}}) \mbox{ and }
    \beta_{c} = \exp{(\cos{(\tilde{\bm{e}}_{i, j}, \bm{e}_{k, l})}}).
\label{eq.5}
\end{equation}

\noindent The total loss for each mini-batch is yielded by:

\begin{equation}
    {\cal L}^{\mbox{\tiny{total}}} = \frac{1}{2N} {\sum_{i=1}^{N}} {{\sum_{j=1}^{2}} {\cal L}_{i,j}^{\mbox{\tiny{W-ACSG}}}}.
\end{equation}
By optimizing the parameters of the speaker encoder with the reformulated contrastive loss, the network is capable of learning robust speaker representations.
\section{Experimental Setup}
\label{sec:pagestyle}

The training set used for training the speaker encoder is Voxceleb2 \cite{chung18b_interspeech}, which contains 1,092,009 utterances from 5,994 speakers. As most of these utterances are continuous speech, we do not apply voice activity detection (VAD) to filter out silent frames. We selected the emphasized channel attention, propagation and aggregation in time-delay neural network (ECAPA-TDNN) \cite{desplanques2020ecapa} as the speaker encoder. We set the channel size to 512. The duration of input audio is 1.8 seconds, and we used the 80-dim log mel-spectrograms extracted from the raw waveforms with a hamming window of 25ms and a frameshift of 10ms as the input features. The output of the speaker encoder is a 192-dim speaker embedding. During the training process, no speaker labels were used.

We evaluated the performance of the speaker encoder on the Voxceleb1 original test set \cite{nagrani17_interspeech}, which consists of 4,874 utterances. We optimized the network with an Adam optimizer. During the testing stage, we calculated the cosine similarity between speaker embeddings of test pairs. The performance metrics are equal error rate (EER) and minimum detection cost function (MinDCF). We set the contrastive learning temperature $\tau$ to 1. The speaker embeddings were L2-normalized. 

To simulate different environments and reduce the effect of non-speaker information, we augmented two non-overlapping segments by adding noise, music, and reverberation randomly sampled from the MUSAN \cite{musan2015} and RIR \cite{ko2017study} datasets. The noise types included ambient noise, television, and babble noise, with SNR ranging from 5 to 20 dB. For the reverberation, we performed a convolution operation with simulated room impulse responses. We removed the discriminator training and followed the same setting as in \cite{tao2022self} to obtain the baseline result. The initial learning rate was 0.001 and it decreased 5\% for every 5 epochs. The batch size was set to 256.

\section{Results and Analysis}
\subsection{Comparison with Existing Works}

We obtained the baseline performance by directly feeding two speaker embeddings into the vanilla contrastive loss (Eq.~\ref{eq.2}). This approach obtained an EER of 8.67\% and a minDCF of 0.50. We compared the performance of the proposed loss function with other state-of-the-art SSL methods in Table~\ref{tab:my_label}. Notably, the proposed method, W-ACSG, outperforms many previous approaches. For instance, \cite{nagrani2020disentangled} employed a cross-modal to provide self-supervision signals for speaker verification. Using a VGG-based network, they achieved promising results by adopting a large batch size. In the case of SimCLR, a larger batch size can potentially yield better performance, due to the presence of numerous negative pairs. As highlighted in \cite{he2020momentum}, negative pairs played an important role in preventing model collapse; as a result, increasing their number helps contrastive learning.

\begin{table}[]
    \centering
    \resizebox{.95\columnwidth}{!}{
    \begin{tabular}{c|c|c|c|c}
         \hline
         Method & Encoder & Batch Size & EER(\%) & minDCF \\
         \hline
         Baseline \tablefootnote{Followed the setting of Tao \textit{et al.} \cite{tao2022self} but removing the discriminator.} & ECAPA-TDNN & 256 & 8.67 & 0.50 \\
         Nagrani et al. \cite{nagrani2020disentangled} & VGG-M & 900 & 22.09 & - \\
         Keoage at al. \cite{lepage2022label} & Thin ResNet-34 & 256 & 13.46 & 0.85 \\
         Huh et al. \cite{huh2020augmentation} & Fast ResNet-34 & 200 & 8.65 & 0.45 \\
         Zhang et al. \cite{zhang2021contrastive} & Thin ResNet-34 & 256 & 8.28 & 0.61 \\
         Xia et al. \cite{xia2021self} & X-vector & 4096 & 8.23 & 0.59 \\
         Mun et al. \cite{mun2020unsupervised} & Fast ResNet-34 & 200 & 8.01 & - \\
         Tao et al. \cite{tao2022self} & ECAPA-TDNN & 256 & 7.36 & - \\
         \hline
         W-ACSG (ours) & ECAPA-TDNN & 256 & \textbf{7.02} & \textbf{0.39} \\ 
         \hline
     
    \end{tabular}}
    
    \caption{Performance comparison of existing and proposed methods on the Voxceleb1 evaluation set. All of the encoders do not have a classification head, and they were trained without using speaker labels.}
    \label{tab:my_label}
\end{table}

Previous works by Huh \textit{et al}. \cite{huh2020augmentation} and Tao \textit{et al}. \cite{ tao2022self} incorporated a discriminator alongside a speaker encoder to suppress the channel effects. However, tuning a discriminator is difficult and not always stable. On the other hand, the MoCo-based approach \cite{lepage2022label} utilizes momentum training and does not require a large batch size. Our proposed method surpasses these classical approaches by introducing clean segments and a weighting mechanism.

\begin{table}[]
    \centering
    \resizebox{.97\columnwidth}{!}{
    \begin{tabular}{c|c|c|c}
         \hline
         Method & Encoder & EER(\%) & minDCF \\
         \hline
         W-ACSG (ours) & \multirow{4}*{ECAPA-TDNN} & 7.02 & 0.39 \\
         w/o Clean segments & & 7.96 & 0.47 \\
         w/o Weighting && 8.04 & 0.45 \\
         w/o both && 8.67 & 0.50 \\
         \hline
    \end{tabular}}
    \caption{The effect of clean segments and weighting mechanism on SV performance.}
    \label{tab:my_label2}
\end{table}
Among the related works, \cite{zhang2021contrastive} is the most similar to ours, as they also incorporated clean segments into a channel-invariant loss. However, their method minimized the Euclidean distance between clean and noisy segments, which was suboptimal in the embedding space when the speaker embeddings were pre-normalized. In our experiments, we only introduced one clean segment for each utterance but achieved better performance.

\subsection{Ablation Study}

We investigated the importance of clean segments and hard negative pairs individually. Table~\ref{tab:my_label2} shows that both methods contribute to the performance gains. When we included clean segments in the training pipeline, the EER was reduced from 8.67\% to 7.96\%. Removing the weighting mechanism increases the EER from 7.06\% to 8.04\%. Combining both methods results in the best performance.
\begin{table}[h]
    \centering
    \resizebox{.52\columnwidth}{!}{
    \begin{tabular}{c|c|c}
         \hline
         Batch Size & EER(\%) & minDCF \\
         \hline
         64 & 8.06 & 0.47 \\
         128 & 7.57 & 0.42 \\
         256 & 7.02 & 0.39\\
         \hline
    \end{tabular}}
    \caption{Ablation study on batch size.}
    \label{tab:my_label3}
\end{table}

Table~\ref{tab:my_label3} shows the effect of batch size on performance. The results show that our system with a mini-batch of 256 achieves the best performance, which closely aligns with the findings in the contrastive learning literature. Our methods surpass many previous works, even with a batch size of 128.

\section{Conclusions}
\label{sec:typestyle}

This paper has explored the impact of clean samples in contrastive learning for speaker verification. We introduced a weighted contrastive loss within an asymmetric self-supervised learning framework for robust speaker representation learning. The proposed method incorporated clean segments to mitigate speaker-dependent information loss. To produce robust speaker embeddings, we optimized the encoder with a weighted loss function. The experimental results on Voxceleb1 demonstrated the merit of the framework.

\footnotesize
\bibliographystyle{IEEEbib}
\bibliography{strings,refs}

\begin{thebibliography}{10}

\bibitem{KINNUNEN201012}
Tomi Kinnunen and Haizhou Li,
\newblock ``An overview of text-independent speaker recognition: From features to supervectors,''
\newblock {\em Speech Communication}, vol. 52, pp. 12--40, 2010.

\bibitem{9889705}
Youzhi Tu, Weiwei Lin, and Man-Wai Mak,
\newblock ``A survey on text-dependent and text-independent speaker verification,''
\newblock {\em IEEE Access}, vol. 10, pp. 99038--99049, 2022.

\bibitem{mak2020machine}
Man-Wai Mak and Jen-Tzung Chien,
\newblock {\em Machine learning for speaker recognition},
\newblock Cambridge University Press, 2020.

\bibitem{eskimez2018front}
Sefik~Emre Eskimez, Peter Soufleris, Zhiyao Duan, and Wendi Heinzelman,
\newblock ``Front-end speech enhancement for commercial speaker verification systems,''
\newblock {\em Speech Communication}, vol. 99, pp. 101--113, 2018.

\bibitem{4806209}
Joseph~P. Campbell, Wade Shen, William~M. Campbell, Reva Schwartz, Jean-Francois Bonastre, and Driss Matrouf,
\newblock ``Forensic speaker recognition,''
\newblock {\em IEEE Signal Processing Magazine}, vol. 26, no. 2, pp. 95--103, 2009.

\bibitem{dehak2010front}
Najim Dehak, Patrick~J Kenny, R{\'e}da Dehak, Pierre Dumouchel, and Pierre Ouellet,
\newblock ``Front-end factor analysis for speaker verification,''
\newblock {\em IEEE Transactions on Audio, Speech, and Language Processing}, vol. 19, no. 4, pp. 788--798, 2010.

\bibitem{prince2007probabilistic}
Simon~JD Prince and James~H Elder,
\newblock ``Probabilistic linear discriminant analysis for inferences about identity,''
\newblock in {\em Proc. IEEE 11th International Conference on Computer Vision}, 2007, pp. 1--8.

\bibitem{snyder2018x}
David Snyder, Daniel Garcia-Romero, Gregory Sell, Daniel Povey, and Sanjeev Khudanpur,
\newblock ``X-vectors: Robust {DNN} embeddings for speaker recognition,''
\newblock in {\em Proc. IEEE International Conference on Acoustics, Speech and Signal Processing}, 2018, pp. 5329--5333.

\bibitem{7846260}
David Snyder, Pegah Ghahremani, Daniel Povey, Daniel Garcia-Romero, Yishay Carmiel, and Sanjeev Khudanpur,
\newblock ``Deep neural network-based speaker embeddings for end-to-end speaker verification,''
\newblock in {\em Proc. IEEE Spoken Language Technology Workshop}, 2016, pp. 165--170.

\bibitem{desplanques2020ecapa}
Brecht Desplanques, Jenthe Thienpondt, and Kris Demuynck,
\newblock ``{ECAPA}-{TDNN}: Emphasized channel attention, propagation and aggregation in {TDNN} based speaker verification,''
\newblock in {\em Proc. Interspeech}, 2020, pp. 3830--3834.

\bibitem{chung2019delving}
Joon~Son Chung, Jaesung Huh, and Seongkyu Mun,
\newblock ``Delving into voxceleb: environment invariant speaker recognition,''
\newblock {\em arXiv preprint arXiv:1910.11238}, 2019.

\bibitem{wan2018generalized}
Li~Wan, Quan Wang, Alan Papir, and Ignacio~Lopez Moreno,
\newblock ``Generalized end-to-end loss for speaker verification,''
\newblock in {\em Proc. IEEE International Conference on Acoustics, Speech and Signal Processing}, 2018, pp. 4879--4883.

\bibitem{liu19f_interspeech}
Yi~Liu, Liang He, and Jia Liu,
\newblock ``Large margin softmax loss for speaker verification,''
\newblock in {\em Proc. Interspeech}, 2019, pp. 2873--2877.

\bibitem{huang2018angular}
Zili Huang, Shuai Wang, and Kai Yu,
\newblock ``Angular softmax for short-duration text-independent speaker verification.,''
\newblock in {\em Proc. Interspeech}, 2018, pp. 3623--3627.

\bibitem{zhang2017end}
Chunlei Zhang and Kazuhito Koishida,
\newblock ``End-to-end text-independent speaker verification with triplet loss on short utterances.,''
\newblock in {\em Proc. Interspeech}, 2017, pp. 1487--1491.

\bibitem{chen2020simple}
Ting Chen, Simon Kornblith, Mohammad Norouzi, and Geoffrey Hinton,
\newblock ``A simple framework for contrastive learning of visual representations,''
\newblock in {\em Proc. International Conference on Machine Learning}, 2020, pp. 1597--1607.

\bibitem{he2020momentum}
Kaiming He, Haoqi Fan, Yuxin Wu, Saining Xie, and Ross Girshick,
\newblock ``Momentum contrast for unsupervised visual representation learning,''
\newblock in {\em Proceedings of the IEEE/CVF Conference on Computer Vision and Pattern Recognition}, 2020, pp. 9729--9738.

\bibitem{zhang2021contrastive}
Haoran Zhang, Yuexian Zou, and Helin Wang,
\newblock ``Contrastive self-supervised learning for text-independent speaker verification,''
\newblock in {\em Proc. IEEE International Conference on Acoustics, Speech and Signal Processing}, 2021, pp. 6713--6717.

\bibitem{xia2021self}
Wei Xia, Chunlei Zhang, Chao Weng, Meng Yu, and Dong Yu,
\newblock ``Self-supervised text-independent speaker verification using prototypical momentum contrastive learning,''
\newblock in {\em Proc. IEEE International Conference on Acoustics, Speech and Signal Processing}, 2021, pp. 6723--6727.

\bibitem{mun2020unsupervised}
Sung~Hwan Mun, Woo~Hyun Kang, Min~Hyun Han, and Nam~Soo Kim,
\newblock ``Unsupervised representation learning for speaker recognition via contrastive equilibrium learning,''
\newblock {\em arXiv preprint arXiv:2010.11433}, 2020.

\bibitem{lepage2022label}
Th{\'e}o Lepage and R{\'e}da Dehak,
\newblock ``Label-efficient self-supervised speaker verification with information maximization and contrastive learning,''
\newblock in {\em Proc. Interspeech}, 2022, pp. 4018--4022.

\bibitem{tian2020makes}
Yonglong Tian, Chen Sun, Ben Poole, Dilip Krishnan, Cordelia Schmid, and Phillip Isola,
\newblock ``What makes for good views for contrastive learning?,''
\newblock {\em Advances in Neural Information Processing Systems}, vol. 33, pp. 6827--6839, 2020.

\bibitem{xiao2021what}
Tete Xiao, Xiaolong Wang, Alexei~A Efros, and Trevor Darrell,
\newblock ``What should not be contrastive in contrastive learning,''
\newblock in {\em Proc. International Conference on Learning Representations}, 2021.

\bibitem{huh2020augmentation}
Jaesung Huh, Hee~Soo Heo, Jingu Kang, Shinji Watanabe, and Joon~Son Chung,
\newblock ``Augmentation adversarial training for self-supervised speaker recognition,''
\newblock {\em arXiv preprint arXiv:2007.12085}, 2020.

\bibitem{tao2022self}
Ruijie Tao, Kong~Aik Lee, Rohan~Kumar Das, Ville Hautam{\"a}ki, and Haizhou Li,
\newblock ``Self-supervised speaker recognition with loss-gated learning,''
\newblock in {\em Proc. IEEE International Conference on Acoustics, Speech and Signal Processing}, 2022, pp. 6142--6146.

\bibitem{li2020contrastive}
Longxin Li, Man-Wai Mak, and Jen-Tzung Chien,
\newblock ``Contrastive adversarial domain adaptation networks for speaker recognition,''
\newblock {\em IEEE Transactions on Neural Networks and Learning Systems}, vol. 33, no. 5, pp. 2236--2245, 2022.

\bibitem{robinson2021contrastive}
Joshua~David Robinson, Ching-Yao Chuang, Suvrit Sra, and Stefanie Jegelka,
\newblock ``Contrastive learning with hard negative samples,''
\newblock in {\em Proc. International Conference on Learning Representations}, 2021.

\bibitem{NEURIPS2020_f7cade80}
Yannis Kalantidis, Mert~Bulent Sariyildiz, Noe Pion, Philippe Weinzaepfel, and Diane Larlus,
\newblock ``Hard negative mixing for contrastive learning,''
\newblock in {\em Advances in Neural Information Processing Systems}, 2020, vol.~33, pp. 21798--21809.

\bibitem{li2023discriminative}
Zhe Li, Man-Wai Mak, and Helen Mei-Ling Meng,
\newblock ``Discriminative speaker representation via contrastive learning with class-aware attention in angular space,''
\newblock in {\em Proc. IEEE International Conference on Acoustics, Speech and Signal Processing}, 2023, pp. 1--5.

\bibitem{chung18b_interspeech}
Joon~Son Chung, Arsha Nagrani, and Andrew Zisserman,
\newblock ``Voxceleb2: Deep speaker recognition,''
\newblock in {\em Proc. Interspeech}, 2018, pp. 1086--1090.

\bibitem{nagrani17_interspeech}
Arsha Nagrani, Joon~Son Chung, and Andrew Zisserman,
\newblock ``Voxceleb: A large-scale speaker identification dataset,''
\newblock in {\em Proc. Interspeech}, 2017, pp. 2616--2620.

\bibitem{musan2015}
David Snyder, Guoguo Chen, and Daniel Povey,
\newblock ``{MUSAN}: {A} music, speech, and noise corpus,''
\newblock {\em arXiv preprint arXiv:1510.08484}, 2015.

\bibitem{ko2017study}
Tom Ko, Vijayaditya Peddinti, Daniel Povey, Michael~L Seltzer, and Sanjeev Khudanpur,
\newblock ``A study on data augmentation of reverberant speech for robust speech recognition,''
\newblock in {\em Proc. IEEE International Conference on Acoustics, Speech and Signal Processing}, 2017, pp. 5220--5224.

\bibitem{nagrani2020disentangled}
Arsha Nagrani, Joon~Son Chung, Samuel Albanie, and Andrew Zisserman,
\newblock ``Disentangled speech embeddings using cross-modal self-supervision,''
\newblock in {\em Proc. IEEE International Conference on Acoustics, Speech and Signal Processing}, 2020, pp. 6829--6833.

\end{thebibliography}

\end{document}